\documentclass[journal]{IEEEtran}

\usepackage{graphicx}  
\begin{document}
\thispagestyle{empty}
\title{The Normalized Radial Basis Function Neural Network and its Relation to the Perceptron}
\author{Igor Grabec\\
Faculty of Mechanical Engineering, University of Ljubljana, Slovenia\\igor.grabec@fs.uni-lj.si}
%\thanks{Manuscript received: January 31, 2006; revised:}%
%\thanks{I. Grabec is with the Faculty of Mechanical Engineering, University of Ljubljana, Slovenia (e-mail:igor.grabec@fs.uni-lj.si)}}
%\markboth{Dynamics of Continuous, Discrete and Impulsive Systems,~B, Special~Volume:~Advances~in~Neural~Networks,~2007}{Shell \MakeLowercase{}}
%\specialpapernotice{(Invited Paper)}
\maketitle
\thispagestyle{empty}
\begin{abstract}
The normalized radial basis function neural network emerges in the statistical modeling of natural laws that relate components of multivariate data. The modeling is based on the kernel estimator of the joint probability density function pertaining to given data. From this function a governing law is extracted by the conditional average estimator. The corresponding nonparametric regression represents a normalized radial basis function neural network and can be related with the multi-layer perceptron equation. In this article an exact equivalence of both paradigms is demonstrated for a one-dimensional case with symmetric triangular basis functions. The transformation provides for a simple interpretation of perceptron parameters in terms of statistical samples of multivariate data.
\end{abstract}

\IEEEpeerreviewmaketitle

\section{Introduction}
Multi-layer perceptrons (MLP) have played a central role in the research
of neural networks \cite{ha,he}. Their study began with the nonlinear and adaptive 
response characteristics of neurons, which have brought with them many 
difficulties related to the understanding of the collective properties of MLPs.
Consequently, it was discovered rather late 
that the MLP is a universal approximator of relations between input signals \cite{ha,he,cy}.  
However, supervised training of MLPs by back-propagation of errors is relatively
time-consuming and does not provide a simple interpretation of MLP 
parameters. The inclusion of {\it a priori} information into an MLP is also 
problematic. Many of these problems do not appear in simulations
of radial basis function neural networks (RBFN) \cite{bi}. The structure of the normalized RBFN
stems from the representation of the empirical probability density function of
sensory signals in terms of prototype data and can simply be 
interpreted statistically \cite{gs}.
An optimal description of relations is described in this case by
the conditional average estimator (CA), which represents a general, non-linear
regression and corresponds to a normalized RBFN. 
{\it A priori} information can also be included in this model by 
initialization of prototypes. A learning rule derived from the maximum 
entropy principle describes a self-organized adaptation of neural 
receptive fields \cite{gs,igso,igexp}. 
The separation of input signals into independent and dependent 
variables need not be done before training, as with MLPs, but it can be 
performed when applying a trained network. Because of these convenient 
properties of RBFNs, our aim was to compare both NN paradigms and to 
explore whether RBFN is equivalent to MLP with respect to modeling of mapping relations. Here we demonstrate their exact 
equivalence for a simple one-dimensional case by showing that the mapping 
relation of an RBFN can be directly transformed into that of an MLP, and vice versa. This further 
indicates how MLP parameters can also be statistically interpreted in the case of multivariate data.

\section{Estimation of probability density functions}
The task of both paradigms is the modeling of relations between components of
measured data. We assume that $D$ sensors provide signals
$(s_1,s_2, \ldots ,s_D)$ that comprise a vector
${\bf x}$. The modeling is here based on an estimation of the 
joint probability density function (PDF) of vector ${\bf x}$. 
We assume that information about the probability distribution is
obtained by a repetition of measurements that yield
$N$ independent samples 
$\lbrace{\bf x}_1,{\bf x}_2,\ldots ,{\bf x}_N\rbrace$. 
The PDF is then described by the kernel estimator [6]: 
\begin{equation}
f_e({\bf x}) = {1 \over N} \sum_{n=1}^N w ({\bf x}-{\bf x}_n
,\sigma)
\label{eq:2}
\end{equation}
Here the kernel $w({\bf x}-{\bf x}_n ,\sigma)$ is a smooth approximation of 
the delta function, such as a radially symmetric Gaussian 
function $w({\bf x} ,\sigma) = {\rm const.}~ \exp (- \parallel {\bf x}\parallel ^2
/2\sigma^2 )$. The constant $\sigma$ can be objectively interpreted as the width of a
scattering function describing stochastic fluctuations in the channels of a data  
acquisition system and can be determined by a calibration procedure \cite{gs,igexp,igextr}.

However, in an application the complete PDF need not be
stored; it is sufficient to preserve a set of statistical samples
${\bf x}_n$. In order to obtain a smooth estimator of the 
PDF, the neighboring sample points should be separated in the 
sample space by approximately $\sim \sigma$. 
From this condition one can estimate a proper number $N$ 
of samples \cite{gs,igexp,igextr}. 
In a continuous measurement the number of samples increases 
without limit, and there arises a problem with the finite capacity of 
the memory in which the data are stored. Neural networks are 
composed of finite numbers of memory cells,
and therefore we must assume that the PDF can be 
represented by a finite number $K$ of prototype 
vectors $\lbrace {\bf q}_1 ,{\bf q}_2,\ldots ,{\bf
q}_K\rbrace$ as
\begin{equation}
f_r({\bf x}) = {1 \over K} \sum_{k=1}^K w({\bf x}-{\bf q}_k,\sigma)
\label{eq:3}
 \end{equation}
\noindent 
In the modeling of $f_r$ the prototypes are first initialized 
by $K$ samples: $\{{\bf q}_k = {\bf x}_k~,~ k=1\ldots K\}$, 
which represent {\it a priori} given information. 
These prototypes can be adapted to additional samples ${\bf x}_N$ 
in such a way that the mean-square difference between $f_e$ and $f_r$ 
is minimized. The corresponding rule was derived elsewhere,  
and it describes the self-organized unsupervised learning of neurons, 
each of which contains one prototype ${\bf q}_k$ \cite{igso,igexp}. 

The estimator of the PDF given in Eq.\, \ref{eq:3} can be simply generalized
by assuming that various prototypes are associated with different probabilities and 
receptive fields \cite{igexp}: ${1 / K}\mapsto p_k$ and $\sigma\mapsto \sigma_k$. This substitution yields a generalized model:
\begin{equation}
f_g({\bf x}) = \sum_{k=1}^K p_k \, w({\bf x}-{\bf q}_k,\sigma_k)
\label{eq:3a}
 \end{equation}
\noindent 
However, in this case several advantages of a simple interpretation of the model Eq.\, \ref{eq:3} are lost, which causes problems when analyzing its relation to the perceptron model. Therefore, we further consider the simpler model given in Eq.\, \ref{eq:3}.  

\section{Conditional average} 
In the application of an adapted PDF the information must be extracted 
from prototypes, which generally corresponds to some kind
of statistical estimation. In a typical application there is some
partial information given, for instance the first $i$ components of the 
vector: ${\bf g}=(s_1,s_2,..,s_i, \emptyset)$, while the hidden data, which
have to be estimated, are then represented by the vector 
${\bf h}=(\emptyset,s_{i+1},..,s_D)$ \cite{gs,igextr}.
Here $\emptyset$ denotes the missing part in a truncated vector. As an 
optimal estimator we apply the conditional average,
which can be expressed by prototype 
vectors as \cite{bi,gs,igextr}:
\begin{equation}
\widehat{\bf h} = \sum_{k=1}^K B_k ({\bf g}) {\bf h}_k,
\label{eq:6}
\end{equation} 
where
\begin{equation}
B_ k({\bf g}) = \frac{w ( {\bf g} - {\bf g}_k ,\sigma)}
{\sum_{j=1}^K w({\bf  g} - {\bf g}_j ,\sigma)}.
\label{eq:6a}
\end{equation}
Here the given vector ${\bf g}$ plays the role of the given condition. 
The basis functions $B_k({\bf g})$ are strongly nonlinear and peaked at
the truncated vectors ${\bf g}_k$. 
They represent the measure of similarity
between the given vector ${\bf g}$ and the prototypes ${\bf g}_k$. 

The CA represents
a general non-linear, non-parametric regression, which has already been successfully applied in a variety of fields \cite{gs,igexp,igextr}. It is important that selection
into given and hidden data can be done after training the network, which essentially contributes to the adaptability of the method to various tasks in an application \cite{gs}.

The CA corresponds to a mapping relation ${\bf g} \rightarrow {\bf h}$
that can be realized by a two-layer RBFN \cite{bi}. The first layer consists of
$K$ neurons. The $k$-th neuron obtains the input signal ${\bf g}$ over synapses
described by ${\bf g}_k$ and is excited as described by the radial basis
function $B_k({\bf g})$. The corresponding excitation signal is then transferred
to the neurons of the second layer. The $i$-th neuron of this layer has
synaptic weights $h_{k,i}$ and generates the output $\widehat{h}_i({\bf g})$.

\section{Transition from RBFN to MLP}
\begin{figure}
\centering
\includegraphics[width=2.5in]{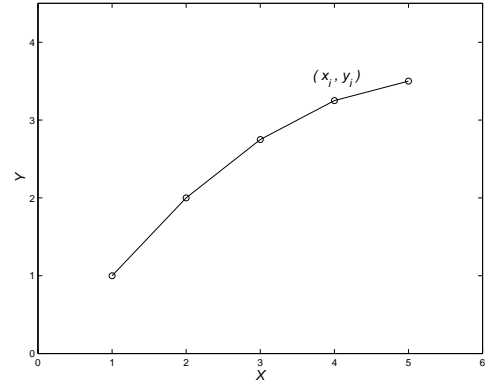}
\caption{An example of a linear interpolating function through sample points.}
\label{fig_1}
\end{figure}

\begin{figure}
\centering
\includegraphics[width=2.5in]{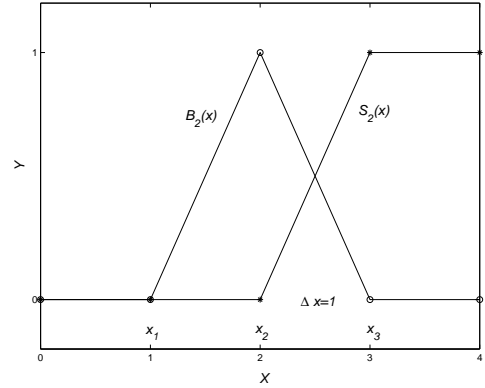}
\caption{Examples of a triangular and a piecewise linear sigmoidal basis functions.}
\label{fig_2}
\end{figure}

In order to obtain a relation with an MLP it is instructive to analyze the
performance of the RBFN in a simple two-dimensional case, for example as shown in Fig. \ref{fig_1}. We consider the
function $y(x)$ described by a set of sample pairs
$\{ x_1, y_1;\ldots x_i, y_i;\ldots ; x_N, y_N\}$ with constant spacing between the sample points:
$\triangle x=x_{j+1}-x_j ~{\rm for}~ j=1\ldots N-1$. We further introduce
a triangular and a piecewise linear sigmoidal basis function, as shown in Fig. \ref{fig_2}:
\begin{eqnarray}
B_i(x) = \lbrace &1&-\quad {\vert x-x_i\vert \over \triangle x}  \quad\ldots \quad  {\rm
for}~ x_{i-1} < x <x_{i+1} ~;~\nonumber \\  
&0&  \ldots  \quad{\rm elsewhere} \quad \rbrace
\label{eq:}
\end{eqnarray}
\begin{eqnarray}
S_i(x)=\{&0&~\ldots \quad{\rm for}~ x<x_i ~;~\nonumber \\ 
&(x-x_i)/\triangle x& ~\ldots\quad{\rm for}
~x_i\leq x\leq x_{i+1}~;~ \nonumber \\
&1&~\ldots\quad{\rm for}~ x > x_{i+1}  \quad\}
\label{eq:7}
\end{eqnarray}
Using these, we can represent the function $y(x)$ by a linear interpolating function comprising straight line segments connecting the sample points. The CA can in this case be readily transformed into an MLP expression
by utilizing the relations:
\begin{equation}
B_{i+1}(x) = S_i (x) - S_{i+1}(x) 
\label{eq:8}
\end{equation}

\begin{equation}
S_i(x) = {B_{i+1}(x)\over B_i(x) + B_{i+1}(x)}
\label{eq:8a}
\end{equation}
The result is:
{\footnotesize
\begin{eqnarray}
\widehat{y}(x)&=&{ y_1 B_1 (x) + \ldots + y_N B_N (x)  \over B_1 (x)+ \ldots + B_N (x)}\nonumber\\ 
&=&{ y_1 B_1 (x)  \over B_1 (x) + \ldots + B_N (x)} +\ldots+ { y_N B_N
(x)  \over B_1 (x) + \ldots + B_N (x)}\nonumber \\ 
&=& { y_1 B_1 (x)  \over B_1 (x) + B_2 (x)} + y_2 B_2 (x) + \ldots \nonumber \\
&& \ldots + y_{N-1} B_{N-1} (x) + { y_N B_N (x)  \over B_{N-1} (x) +B_N (x)}
\label{eq:9}
\end{eqnarray}}
\noindent In the denominator of the first and last terms of this expression,
only those basis functions are kept that differ from zero in the region
where the basis function in the numerator also differs from zero. The denominator
in terms of index $2$ to $N-1$ is $1$ because of the overlapping of
neighboring basis functions. We insert relations of Eq.\,(\ref{eq:8},\ref{eq:8a}) into
Eq.\,(\ref{eq:9}) and obtain
\begin{equation}
\widehat{y}(x)= y_1+ \sum_{i=1}^{N-1} (y_{i+1}-y_i )\; S_i (x)
\label{eq:10}
\end{equation}
By introducing the parameters: $\triangle y_i= y_{i+1}-y_i ,\quad c_i =1/(x_{i+1}-x_i)
,\quad \Theta _i = x_i /(x_{i+1}- x_i )$\, and a unique, normalized sigmoidal basis function:
\begin{eqnarray}
S(x) = \{ &0& \ldots ~{\rm for}~ x<0 ~;~\nonumber\\  
&x& \ldots ~{\rm for}~ 0\leq x\leq 1~ ;~\nonumber\\
&1& \ldots ~{\rm for}~ x>1 \quad \}
\label{eq:11}
\end{eqnarray}
we can write Eq.\,(\ref{eq:10}) in the form of a two-layer perceptron mapping
relation
\begin{equation}
\widehat{y}(x)= y_1 + \sum_{i=1}^{N-1} \triangle y_i \,S(c_i  x -\Theta_i)
\label{eq:12}
\end{equation}
The first layer corresponds to neurons with synaptic weights $c_i$
and threshold values $\Theta_i$, while the second layer contains a linear
neuron with synaptic weights $\triangle y_i$ and threshold $y_1$.

The above derivation demonstrates that for the two-dimensional distribution 
the mapping $x \rightarrow y$ determined by the conditional average is identical
with the mapping relation of a multi-layer perceptron. However, a difference
appears when the operations needed for the mapping are executed. The operators
involved in both cases are described by different basis functions, which
correspond to different neurons in the implementation. If the prototypes
are not evenly spaced, then the last equation can still be applied, although
the transition regions will be of different spans. However, in this case
the basis functions $B_i (x)$ are no longer symmetric. In applications
it is more convenient to use a Gaussian basis function rather than 
a triangular one, and in the perceptron expression this yields the function 
$\tanh (\ldots)$. In this case, the estimated function $\widehat{y}(x)$ generally does not
run through the sample points but rather approximates them by a function
having a more smooth derivative than the piecewise linear function. In this
case, the correspondence between RBFN and MLP is not exact but approximate.

An additional interpretation is needed when the data are not
related by a regular function $y(x)$ but randomly, as described by a joint
probability density function $f(x,y)$. In this case, various values of
$y$ can be observed at a given $x$. Evaluation of CA in this case is not
problematic, while in the perceptron relation Eq.\,(\ref{eq:12}) the value
$y_i$ must be substituted by the conditional average of variable $y$ at $x_i$. 

The analysis of the correspondence between RBFN and MLP can be extended
to multi-variate mappings. Let us first consider the situation with just two
prototypes ${\bf q}_i$ and ${\bf q}_j$ and Gaussian basis functions. 
The CA is then described by the function
\begin{equation}
\widehat{\bf h}({\bf g}) =
{{\bf h}_i\exp ({-  \parallel{\bf g}-  {\bf g}_i\parallel ^2 \over 2 \sigma^2})
+{\bf h}_j\exp ({-  \parallel{\bf g}-  {\bf g}_j\parallel ^2 \over 2 \sigma^2})
 \over
 \exp ({- \parallel {\bf g}-  {\bf g}_i \parallel ^2 \over 2 \sigma^2})
+\exp ({- \parallel {\bf g}-  {\bf g}_j \parallel  ^2 \over 2 \sigma^2})
}
\label{eq:13}
\end{equation}
We introduce the notation:
${\bf g}_i=\overline {\bf g}-\triangle {\bf g}~ ,~
{\bf g}_j=\overline {\bf g}+\triangle {\bf g}~ ,~
{\bf h}_i=\overline {\bf h}-\triangle {\bf h}~ ,~
{\bf h}_j=\overline {\bf h}+\triangle {\bf h}$~
in which the overline denotes the average value and $2\triangle {\bf g}$ is
the spacing of the prototypes. If we express the norm by a scalar product
and cancel the term $\exp[-(\parallel {\bf g}-\overline{\bf g} \parallel
^2+\parallel \triangle {\bf g}\parallel ^2)/2\sigma^2)$ in the numerator
and denominator, we obtain the expression:
\begin{equation}
\widehat{\bf h}({\bf g}) = \overline{\bf h} + \triangle {\bf h} \tanh
\left[\triangle {\bf g}\cdot ({\bf g}- \overline{\bf g})/\sigma^2\right]
\label{eq:14}
\end{equation}
in which $\cdot$  denotes the scalar product. In order to obtain the relation
between RBFN and MLP, we introduce a weight vector ${\bf c} =\triangle
{\bf g}/\sigma^2$ and a threshold value $\Theta = \overline{\bf g}\cdot
\triangle {\bf g}/\sigma^2$ into Eq.\,(\ref{eq:14}) and obtain:
\begin{equation}
\widehat{\bf h}({\bf g}) =  \overline{\bf h} + \triangle {\bf h} \tanh
\left[{\bf c}\cdot ({\bf g}- \overline{\bf g}) - \Theta\right]
\label{eq:15}
\end{equation}
This expression again describes a two-layer perceptron: the first layer
is composed of one neuron having the synaptic weights described by the
vector ${\bf c}$ and the threshold value $\Theta$. The second layer is
composed of linear neurons having synaptic weights $\triangle h_i$ and
threshold values $\overline{h_i}$. 

The first-order approximation of the
mapping expression Eq.\, \ref{eq:15} is\,:
\begin{equation}
\widehat{\bf h}({\bf g}) =  \overline{\bf h} + \triangle {\bf h}\triangle
{\bf g}\cdot ({\bf g}-  \overline{\bf g})/\sigma^2
\label{eq:16}
\end{equation}
This equation represents a linear regression of ${\bf h}$ on ${\bf g}$
that runs through both prototype points if we assign $\sigma^2= \parallel
\triangle {\bf g}\parallel ^2$. Its slope is determined by the covariance
matrix ${\bf \Sigma}=\triangle {\bf h}\triangle {\bf g}^{\rm T}$. However,
the nonlinear regression specified in Eq.\,(\ref{eq:14}) follows a linear
regression only in the vicinity of a point determined by $\overline{\bf
g}$ and $\overline{\bf h}$ while it exhibits saturation when ${\bf g}$
runs from $\overline{\bf g}$ over given prototypes to infinity. The saturation
is a consequence of the function $\tanh (\ldots)$, which is basic in the modeling
of a multi-layered perceptron.

The reasoning presented above for a multi-variate case requires additional
explanation when transferred to a situation consisting of many prototypes.
Let us assume that $N$ prototypes with indexes $1 \ldots N$ can be found
in the hyper-sphere of radius approximately $\sigma$ around the given
datum ${\bf g}$, and let these prototypes be spaced by approximately equal
distances. The CA can now be expressed with leading terms and remainders as follows\,:
\begin{equation}
\widehat{\bf h} ({\bf g}) = {
\sum_{i=1}^N{\bf h}_i \exp (-\parallel {\bf g}- {\bf g}_i\parallel^2
/2\sigma^2)\over \sum_{i=1}^N \exp (-\parallel {\bf g}- 
{\bf g}_i\parallel^2 /2\sigma^2)+O_w} +O_h 
\label{eq:17}
\end{equation} 
Here $O_h$ and $O_w$ represent two remainders, which are small in comparison
with the two leading terms. We again introduce the average value, but
now with respect to $N$ prototypes: ${\bf g}_i= \overline{\bf g}+\triangle
{\bf g}_i ~,~ {\bf h}_i = \overline{\bf h}+\triangle {\bf h}_i ~{\rm for}~
i=1 \ldots N$. With this we obtain the approximate expression\,:
\begin{equation}
\widehat{\bf h}({\bf g}) \cong \overline{\bf h}+
{\sum_{i=1}^N \triangle {\bf h}_i \exp [\triangle {\bf g}_i \cdot ({\bf
g}-\overline{\bf g})/\sigma^2]  \over
\sum_{i=1}^N \exp [\triangle {\bf g}_i \cdot ({\bf g}-\overline{\bf
g})/\sigma^2]}
\label{eq:18}
\end{equation}
For ${\bf g}$ in the vicinity of the average value, a linear approximation
of the exponential function is applicable, which yields
\begin{equation}
\widehat{\bf h}({\bf g}) \cong \overline{\bf h}+{1\over N}\sum_{i=1}^N
\triangle {\bf h}_i \triangle {\bf g}_i \cdot ({\bf g}- \overline{\bf
g})/\sigma^2
\label{eq:19}
\end{equation}
This expression represents a linear regression of ${\bf h}$ on ${\bf g}$
specified by $N$ points. If we express the covariance matrix 
\begin{equation}
{\bf \Sigma} = {1\over N}\sum_{i=1}^N\triangle {\bf h}_i \triangle {\bf
g}_i^{\rm T} 
\label{eq:20a}
\end{equation}
by two principal
vectors $\triangle {\bf h}_p$ and $\triangle {\bf g}_p$ :
\begin{equation}
{\bf \Sigma} = \triangle {\bf h}_p \triangle {\bf g}_p^{\rm T}
\label{eq:20}
\end{equation}
we obtain a simplified expression of the linear regression
\begin{equation}
\widehat{\bf h}({\bf g}) \cong \overline{\bf h} + \triangle {\bf h}_p
\triangle {\bf g}_p \cdot ({\bf g}- \overline{\bf g})/\sigma^2
\label{eq:21}
\end{equation}
which is an approximation of an MLP mapping relation
\begin{equation}
\widehat{\bf h}({\bf g}) \cong \overline{\bf h} + \triangle {\bf h}_p
\tanh [\triangle {\bf g}_p\cdot ({\bf g}-\overline{\bf g})/\sigma^2]
\label{eq:22}
\end{equation}
The parameters of a single neuron in the perceptron expression thus correspond
to the principal vectors of the covariance matrix 
${\bf \Sigma}=\triangle {\bf h}_p \triangle {\bf g}_p^{\rm T}$ determining a local
regression around the center of several neighboring prototypes. 

The above
expression shows that the transition from RBFN to MLP can be quite generally
performed. However, in the multi-variate case, the decomposition of CA
into a perceptron mapping is not as simple as in the one-dimensional case,
because the interpretation of perceptron parameters goes over local regression
determined by various prototypes surrounding the given datum ${\bf g}$. In spite of this, our conjecture is that both paradigms are equivalent with respect to the statistical modeling of mapping relations, provided that both models include the same number of adaptable parameters. 

\section{Conclusion}
The conditional average 
representing a linear interpolating function by the regular function $y(x)$ shown in Fig.\,\ref{fig_1} can be exactly decomposed into the multilayer perceptron relation. When there are a small number of noise-corrupted sample data points representing the function, the question of proper smoothing arises. In the case of CA this is done by using symmetric radial basis functions and increasing their width. The basis functions centered at various points then overlap, which results in a smoother
$\widehat{y}(x)$. Because of multiple overlapping, the relations between
radial basis and sigmoidal functions becomes more complicated, and the transition
between the conditional average and the perceptron relation becomes less obvious.
However, when the prototypes are obtained by self-organization, they represent
a statistical regularity, and the CA generally does not exhibit statistical
fluctuations. In this case, the proper RBF width corresponds to the
distance between closest neighbors, and additional smoothing is not needed. 
The corresponding parameters of the perceptron for one-dimensional mapping can then be simply interpreted in terms of prototypes, as described by the model equations Eq. \ref{eq:6} and Eq. \ref{eq:22}. However, due to the complexity of the self-organized formation of prototypes determining the RBFN and the back-propagation learning of the MLP, it would be difficult to find an exact mapping relation between both models, especially in the multivariate case. 

\section*{Acknowledgment}
% optional entry into table of contents (if used)
%\addcontentsline{toc}{section}{Acknowledgment}
This work was supported by the Ministry of Higher Education, Science and Technology of the Republic of Slovenia and EU-COST.
The author thanks Prof. W. Sachse from Cornell University, Ithaca, NY, USA for his valuable suggestions in the preparation of this article.


\noindent

\begin{thebibliography}{1}
\bibitem{ha} S.~Haykin, \emph{Neural Networks, A Comprehensive Foundation}, 2nd ed.\hskip 1em plus 0.5em minus 0.4em\relax   New York, NY: Macmillan, 1999.

\bibitem{he} R.~Hecht-Nielsen, \emph{Neurocomputing},\hskip 1em plus 0.5em minus 0.4em\relax   Reading, MA: Addison-Wesley, 1990.

\bibitem{cy} G.~Cybenko, "Approximations by Superpositions of a Sigmoidal Function," \emph{Math. Cont., Sig. \& Syst.}, vol. 2, pp. 303-314, 1989. 

\bibitem{bi} C.~M.~Bishop, "Neural Networks and their Applications," \emph{Rev. Sci. Instr.}, vol. 65, pp. 1830-1832, 1994. 

\bibitem{gs} I.~Grabec and W.~Sachse, \emph{Synergetics of Measurement, Prediction and Control}, Berlin: Springer-Verlag, 1997.

\bibitem{igso} I.~Grabec, "Self-Organization of Neurons Described by the Maximum Entropy Principle," \emph{Biological Cybernetics}, vol. 69~(9), pp. 403-409, 1990.

\bibitem{igexp} I.~Grabec, "Experimental Modeling of Physical Laws," \emph{Eur. Phys. J. B}, vol. 22, pp. 129-135, 2001.

\bibitem{igextr} I.~Grabec, "Extraction of Physical Laws from Joint Experimental Data," \emph{Eur. Phys. J. B}, vol. 48, pp. 279-289, 2005.

\bibitem{dh} R.~O.~Duda and P.~E.~Hart, \emph{Pattern Classification and Scene Analysis},\hskip 1em plus 0.5em minus 0.4em\relax  New York: J. Wiley and Sons, 1973, Ch. 4.

\end{thebibliography}
\end{document}